\documentclass[aps,prc,preprint,showpacs]{revtex4}
\usepackage{graphicx}
\begin{document}
\bibliographystyle{num}
\title{Dynamical response functions in correlated fermionic systems}
\baselineskip=1. \baselineskip

\author{     P. Bo\.{z}ek\footnote{Electronic
address~:
piotr.bozek@ifj.edu.pl}}
\affiliation{
Institute of Nuclear Physics, PL-31-342 Cracow, Poland}
\author{
J. Margueron and H. M\"uther}
\affiliation{
Institut f\"ur Theoretische Physik, Universit\"at T\"ubingen, Germany}

\date{\today}




\begin{abstract}
Response functions in nuclear matter at finite temperature are considered
beyond the usual Hartree-Fock (HF) plus Random Phase Approximation (RPA) scheme.
The contributions  due to the propagator for the dressed nucleons and the
corresponding vertex corrections are treated in a consistent way. For that
purpose a semi-realistic Hamiltonian is developed with parameters adjusted to
reproduce the nucleon self-energy as derived from realistic nucleon-nucleon
interactions.  For a scalar residual interaction the resulting response
functions are very close to the RPA response functions. However, the collective
modes, if present, get an additional width due to the coupling to multi-pair
configurations. For  isospin dependent residual interactions we find strong
modifications of isospin response functions due to multi-pair contributions in
the response function.  Such a modification can lead to the disappearance of
collective spin or isospin modes in a correlated system and shall have an
effect on the absorption rate of neutrinos in nuclear matter.
\end{abstract}


\pacs
{\bf  21.65.+f, 24.10.Cn, 26.60.+c}

\maketitle


\section{Introduction}
The shell-model or independent particle model (IPM) has been very successful
in describing basic features of nuclear systems. This means that nuclei are
considered as a system of nucleons moving independently in a mean field and
the  residual interaction between these particle or quasiparticles is supposed
to be weak. Therefore the response of the the system to an external
perturbation can be calculated within the Fermi Liquid theory \cite{pines} in
terms of linear response functions. These response functions are calculated
assuming a Hartree-Fock (HF) propagator for the particle-hole excitations of
the nucleons and including the residual interaction by means of the Random
Phase Approximation (RPA) approximation. In the long-wavelength limit or
external  perturbations with low momentum transfer, the residual interaction
between the quasiparticles is usually parameterized in terms of Landau
parameters.

This HF plus RPA scheme is typically used to determine e.g. the neutrino
propagator in hot and dense nuclear
matter \cite{mf,reddyne,voskrn,yamada,raffelt,sedr} and it has been found
that the neutrino opacity is very sensitive to the details of these
response functions.   This quantity is very crucial for the simulation of
astrophysical objects like the explosion of supernovae or the cooling mechanism
for neutron stars \cite{haensel,jerom1}. 

The study of the response is also very important to determine the propagator of
mesons or a photon in the nuclear medium. Therefore such investigations have to
be performed to explore e.g. the possibility for a pion
condensation\cite{pico1,pico2} or the production and emission of mesons and
photons from the hot dense matter obtained in heavy ion reactions
\cite{Bozek:1997rv,Bozek:1998ro,eff,Cassing:2000ch,Bertsch:1996ig,Knoll:1996nz}. Last not
least the response function is also reflecting the excitation modes of nuclei.

However, the simple HF plus RPA scheme outlined above is applicable to nuclear
systems only if effective nucleon-nucleon (NN) interactions like
Skyrme\cite{skyrme} or Gogny\cite{gogny} forces are employed. The IPM fails
completely if realistic  NN interactions are considered, which have been
adjusted to describe the NN scattering data. Trying to evaluate the energy of
nuclear matter from such realistic interaction within the HF approximation
yields positive energies, i.e. unbound nuclei\cite{polrep}. The reason of this
deficiency of the HF approximation in nuclear physics are the correlations
beyond the IPM approach, which are induced from the strong short-range and
tensor components of a realistic NN interaction.

These correlations have a significant effect on the single-particle propagator
for a nucleon in the nuclear medium. The spectral function still exhibits a
quasiparticle peak. A sizable fraction, however, of the strength occurs at
energies above and below the quasiparticle peak. For hole-states one typically
observes that around 15 \% of the spectral strength is shifted to energies above
the Fermi energy \cite{Bozek:2002em,Dewulf:2003nj,Frick:2003sd} which means that
the occupation probability of those states is reduced from 100 \% in the case of
the IPM approach to around 85 \%. Another fraction of the hole-strength is 
shifted to energies below the quasiparticle energies, which means that it should
be found in nucleon knock-out experiments at large missing energies. These
effects of correlations on the spectral distribution are confirmed in
($e,e'p$) experiments (see e.g.\cite{rohe:2004}).

In lowest order this redistribution of the strength in the single-particle 
spectral function is due to the admixture of two-hole one-particle and 
two-particle one-hole contributions to the propagator in the HF field. Therefore
one may feel the temptation to use these correlated propagators and evaluate the
nuclear response function in terms of these dressed propagators. In this way one
is including two-particle two-hole admixtures to the particle-hole response
function. As we can expect from the discussion above and as we will see below,
such a procedure leads to response functions which, comparing to the HF plus RPA
response, exhibit a significant shift of the excitation strength to larger
energies.

We will also see, however, that such a significant shift of the excitation
strength, can in general not be consistent with the energy weighted sum-rules,
which are observed in the HF plus RPA scheme. It is well known that it is
rather difficult to develop a symmetry conserving approach for the evaluation
of Green's functions which accounts for correlations beyond the HF plus RPA
approximation. In the case of the response function this requires a consistent
treatment of propagator and vertex corrections.  In this manuscript we will
follow the general recipe of Baym and Kadanoff\cite{kadBaym,blarip} for
calculating the in-medium coupling of an external perturbation to dressed
nucleons in a self-consistent way.

This procedure leads to an integral equation, a Bethe-Salpeter equation for the
dressed vertex. The in-medium vertex has the structure of a three-point Green's
functions. For dressed (off-shell) nucleons it is a function of two momenta and
two energies. Since such calculations are rather involved, only a few exist for
the density response function \cite{bonitz,faleev,pbpl}. In order to make these
calculation feasible, we define a simple interaction model.  The nucleons are
dressed by a  mean-field and a residual interaction. The residual interaction
is taken selfconsistently to the second order. The parameters of the
interaction are adjusted to reproduce the main features of single-particle
spectral functions derived from realistic NN interactions. 

After this introduction we present in section 2 our interaction model and
the adjustment of its parameters to reproduce the nucleon self-energy derived
from a realistic interaction. The evaluation of the response functions with a
consistent treatment of propagator and vertex corrections is outlined in section
3. Numerical results for symmetric nuclear matter and pure neutron matter are
presented in section 4.  There we also discuss the effect of multi-pair
contributions to the response functions at high excitation energies and its
relation to the spin-isospin structure of the residual interaction and the
consequences for the damping of collective modes. The final section summarizes
the main conclusions of this study.   

\section{Mean-field and residual interaction}

\label{residualmf}

Calculations of the response function in nuclear matter are usually
restricted to the HF plus RPA approximation, employing parameterizations
of the effective NN interaction. 
There exist many relatively simple and successful parameterization of
the mean-field Hamiltonian for nuclear systems, e.g. The Skyrme
interaction \cite{skyrme} and the Gogny interaction \cite{gogny}.
Usually such effective interactions include spin and isospin dependent
terms, and also density dependent terms.
The Skyrme interaction is  a zero range interaction with velocity
dependent terms, for which a complete calculation of the RPA response is
possible
\cite{GNVS}. In general, however, the calculation of  the RPA response requires 
the consideration of a nontrivial sum of exchange terms\cite{pico1,depace},
which are often approximated.
Usually the response function
for finite range interactions is calculated expanding the interaction
in Landau parameters \cite{jerom2}.

\begin{figure}
\centering
\includegraphics*[width=0.7\textwidth]{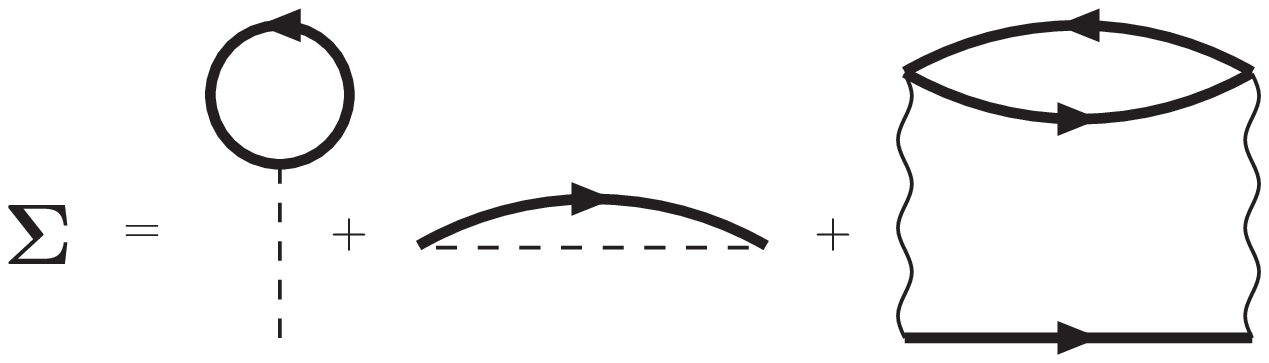} 
\parbox{14cm}{\it
\caption{  Diagrams for the self-energy. The first two diagrams
  are the Hartree-Fock contribution for the Gogny interaction (the dashed
  line). The last diagram is the contribution of the residual interaction in 
  the second order.}
\label{selffig}}
\end{figure}

In this study we want to analyze the linear response functions in a
fermionic systems when the correlations of the system
that  we take into account go beyond the mean-field approximation. 
We suppose that the 
 interactions between the nucleons are given by a mean-field
potential and a residual interaction. The mean-field interaction that we
take is based on the Gogny parameterization
\begin{equation}
\label{Vgogny}
V_{mf}(1,2)=\sum_{i}\left(W_i+B_iP^\sigma-H_iP\tau-M_i P^\sigma P^\tau\right)
e^{-(\bf{r}_1-\bf{r}_2)^2/\mu_i^2} 
+ \sum_{j}t^j_3(1+x_3^j
P^\sigma)\rho^{\sigma_j}\delta^3(\bf{r}_1-\bf{r}_2) \ .
\end{equation}
 The first term is a sum of two Gaussians giving a
finite range interaction and the second term is a sum of two zero-range
density dependent interactions
\cite{gognyd1p}, $P^\sigma=\frac{1}{2}(1+\sigma_1 
 \sigma_2)$ and $P^\tau=\frac{1}{2}(1+\tau_1 \tau_2)$.
 The residual interaction is taken in a very simple form \cite{dan}
\begin{equation}
\label{vres}
V_{res}({\bf r_1}-{\bf r_2})=V_0 e^{-({\bf r_1}-{\bf r_2})^2/2\eta^2}\ ,
\end{equation}
with the parameters
$V_0=453$MeV, $\eta=0.57$fm. The single particle propagator is
calculated by taking the mean field contributions only for the 
Gogny interaction (\ref{Vgogny}) and the second order direct Born term for the
residual interaction (\ref{vres}). The relevant diagrams are shown in
Fig.
\ref{selffig}.

\begin{figure}
\centering
\includegraphics*[width=0.5\textwidth]{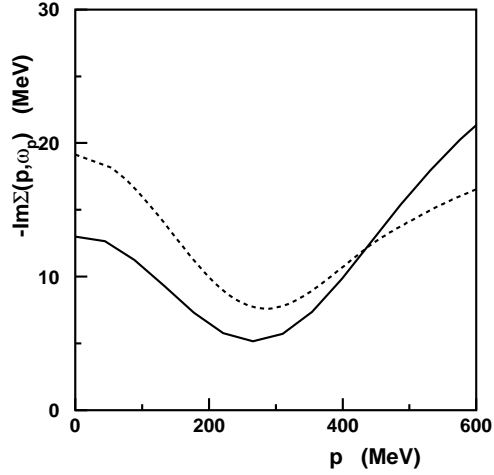}
\parbox{14cm}{\it\caption{Imaginary part of the self-energy at the
quasi-particle pole from the residual interaction (solid line) and from a
self-consistent  $T$-matrix calculation \cite{bcapp} (dashed line). The curves
shown correspond to symmetric nuclear matter at normal nuclear density and
temperature $T=15$MeV.}
\label{szer}}
\end{figure}

The residual interaction induces a finite width to the nucleon
excitations in the medium. Such a dressing of nucleons is expected in
any approach going beyond the simple mean-field. Calculation for the
system
including the mean-field and residual interactions are performed 
in the real-time representation for the thermal Green's functions 
\cite{keldysh}.
The iterated system of equations includes expressions for the
self-energies,
\begin{eqnarray}
\label{self}
\Sigma_{mf}(p)&=&V_{mf}(0)\rho - Tr \int \frac{d^3k}{(2\pi)^3}
P^\sigma P^\tau V_{mf}({\bf p}-{\bf k}) n(k) \nonumber \ , \\
\Sigma^{>(<)}({\bf p},\omega)& =& 4 i \int 
\frac{d^3p_1 d\omega_1 d^3p_2 d\omega_2}
{(2 \pi)^8}
V^2_{res}({\bf p}-{\bf p_1})G^{>(<)}({\bf p_1},\omega_1)
G^{<(>)}({\bf p_2},\omega_2) \nonumber \\
 & & G^{>(<)}({\bf p}-{\bf p_1}+{\bf p_2},\omega-\omega_1+\omega_2) \ ,
\end{eqnarray}
\begin{equation}
\label{dispself}
\Sigma^{r(a)}({\bf p},\omega)=\Sigma_{mf}(p)+
\int\frac{d\omega_1}{2\pi}\frac{\Sigma^<({\bf p},\omega_1)
-\Sigma^>({\bf p},\omega_1)}{\omega-\omega_1\pm i\epsilon} \ ,
\end{equation}
and the Dyson
equation for the retarded (advanced) Green's functions
\begin{equation}
\label{dyson}
G^{r(a)}({\bf p},\omega)=\frac{1}{\omega-{\bf p}^2/2m
-\Sigma^{r(a)}({\bf p},\omega)} \ .
\end{equation}
The Green's functions
\begin{eqnarray}
G^{>}({\bf p},\omega) &= & -i \left(1-f(\omega)\right)A({\bf
 p},\omega) \ ,
 \nonumber \\
G^{<}({\bf p},\omega) &= & i f(\omega)A({\bf p},\omega) 
\end{eqnarray}
are written using the Fermi distribution $f(\omega)$ and
the spectral function
\begin{equation}
\label{spec}
A({\bf p},\omega)=-2 {\rm Im} G^r({\bf p},\omega) \ .
\end{equation}
The nucleon momentum distribution is 
\begin{equation}
n(p)=\int \frac{d\omega}{2\pi} A(p,\omega)f(\omega) \ 
\end{equation}
and the  chemical potential is adjusted at each iteration 
to reproduce the assumed density
\begin{equation}
\rho=4 \int\frac{d^3p}{(2\pi)^3}n(p) \ .
\end{equation}
The  self-consistent equations for the 
one-body properties have been   solved within similar approximations  by
several groups 
\cite{dan,oset,Bozek:1998ro,eff,Lehr:2000ua}.
The width obtained from a self-consistent calculation in the second
order of the residual interaction is similar to the result obtained
from a self-consistent $T$-matrix calculation using realistic bare
nucleon-nucleon interaction (Fig. \ref{szer}).

\begin{figure}
\centering
\includegraphics*[width=0.5\textwidth]{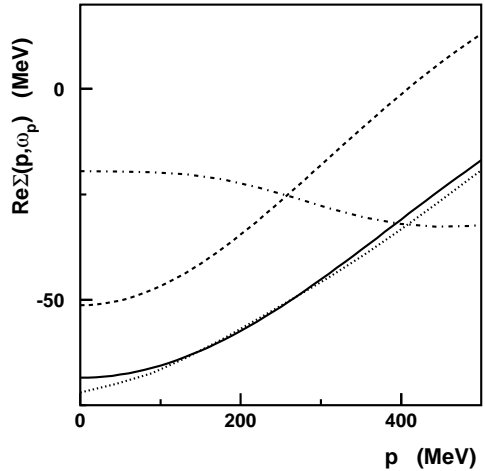}
\parbox{14cm}{\it\caption{Real part of the self-energy at the quasi-particle
pole from the residual interaction (dashed-dotted line) and from the modified
mean-field interaction (dashed line). The sum of the  two contributions is
shown as the solid line and compared the original Gogny potential (dotted
line).}
\label{realsf}}
\end{figure}

The real part of the self-energy $\mbox{Re }\Sigma^r(p,\omega)$ is the sum of
the mean-field  (Gogny) contribution and a dispersive one obtained
from the dispersion relation in Eq. (\ref{dispself}). This means that
the  real part of the self-energy at the quasiparticle pole
$\omega_p=p^2/2m+ \Sigma(p,\omega_p)$
is different from the Gogny single-particle potential.
Accordingly we have modified some parameters of the Gogny interaction
in order to have the same Fermi energy as function of density and a
similar effective mass.
In Fig. \ref{realsf} we show the real part of the self-energy at the
quasi-particle pole and compare it to the single-particle potential derived from
the original Gogny interaction.
This real part of the self-energy is the sum of the dispersive part and
the mean-field contribution originating from the modified Gogny interaction,
which are also shown.

The parameters of the mean-field interaction are given in Table
\ref{table}. We modify the  parameters $M_i$ to reproduce the momentum
dependence of the self-energy and also the density dependent zero
range term, an additional term of the form $t_3^0  \delta^3({\bf
  r}_1-{\bf r}_2)$ is added to the mean-field interaction.
\begin{table}
\begin{tabular}{|r|c|c|c|}
\hline
parameter & Gogny $D1P$ & Modif. I & Modif II \\
\hline
$\mu_1$ (fm) & 0.9 & - & - \\
$\mu_2$ (fm) & 1.44 & - & - \\
$W_1$ (MeV) & -372.9 & - & - \\
$W_2$ (MeV) & 34.6 & - & - \\
$B_1$ (MeV) & 62.7 & - & - \\
$B_2$ (MeV) & -14.1 & - & - \\
$H_1$ (MeV) & -464.5 & - & - \\
$H_2$ (MeV) & -70.9 & - & - \\
$M_1$ (MeV) & -31.5 &  38.5 &  38.5\\
$M_2$ (MeV) & -21 & -51 & -51  \\
$\sigma_1$ & .33 & - & -\\
$\sigma_2$ & .92 & - & -\\
$ t_3^1$ (MeV fm$ ^{3(\sigma_1+1)}$) & 1025.9 & 454.7 & -245.3 \\
$ t_3^2$  (MeV fm$ ^{3(\sigma_2+1)}$)& 1025.9 & - & - \\
$ t_3^1 x_3^1$ (MeV fm$ ^{3(\sigma_1+1)}$) & 1190 & - & -\\
$ t_3^2 x_3^2$ (MeV fm$ ^{3(\sigma_2+1)}$)  & 256 & - & -\\
$ t_3^0$ (MeV fm$ ^{3}$)  & 0 & 478.4 & 803.4 \\
\hline
\end{tabular}
\parbox{14cm}{\it\caption{Table of  the parameters for the mean-field
interaction. The first column corresponds to the Gogny $D1P$ parameterization
\cite{gognyd1p}, the second and third columns are modifications of the
mean-field interaction  used in symmetric nuclear and neutron matter,
respectively. A dash is put whenever the value of the corresponding parameter
is not changed.}\label{table}}
\end{table}

\section{Correlations and response functions}

\label{respsect}

In this section we discuss
 the linear
 response of a correlated system to an external perturbation. As it has already
been mentioned above the evaluation of the RPA response function requires for a
general interaction a solution of a Bethe-Salpeter (BS) equation with a
non-trivial kernel which is due to the exchange terms in the NN interaction.
Therefore one often simplifies the solution of the BS equation by a
parameterization of the particle-hole interaction in terms of 
Landau parameters.
E.g. the  density response function is written using the zero order Landau
parameter $f_0$ as
\begin{equation}
\Pi^r(p,\omega)=\frac{\Pi^r_0(p,\omega)}{1-f_0 \Pi_0^r(p,\omega)} \ ,
\end{equation}
where $\Pi_0^r(p,\omega)$ is the response function of the free Fermi
gas using an effective mass to describe the momentum dependence of the mean
field.

\begin{figure}
\centering
\includegraphics*[width=0.6\textwidth]{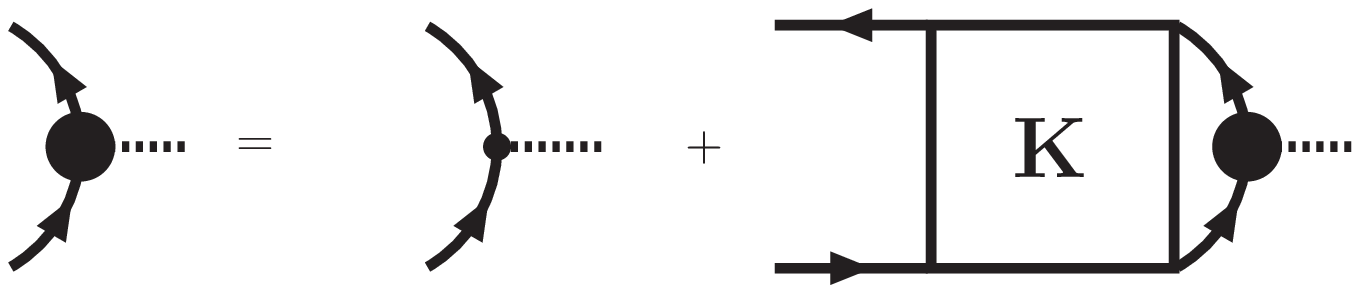}
\parbox{14cm}{\it\caption{The Bethe-Salpeter equation for the dressed vertex.
The particle-hole irreducible kernel $K$ is denoted by the box and the fat and
the small dots denote  the dressed and the bare vertices for the coupling of
the external field to the nucleon. All the fermionic propagators are dressed by
the self-energy as displayed in Fig. \ref{selffig}.}
\label{bsfig}}
\end{figure}

When the description of the correlated systems goes beyond the
mean-field approximation  the difficulty involved in  a 
consistent calculation of the response
function is severely increased. A naive calculation of the 
 polarization bubble using dressed propagators
\begin{equation}
\label{polsimp}
\Pi^{<(>)}({\bf q},\Omega)=- 4 i  \int\frac{d^3p d\omega}{(2\pi)^4}
G^{<(>)}({\bf q}+{\bf p},\omega+\Omega)G^{>(<)}({\bf p},\omega) \ 
\end{equation}
and
\begin{equation}
\label{disppol}
\Pi^{r(a)}(p,\omega)=\int\frac{d\omega_1}{2\pi}\frac{\Pi^<({ p},\omega_1)
-\Pi^>({ p},\omega_1)}{\omega-\omega_1\pm i\epsilon} \ ,
\end{equation}
can be a rather poor estimate for the response function \cite{pbpl}. In
 particular, it
severely violates the $\omega$-sum rule for finite momentum $q$.
This violation of the sum rule by the naive one-loop response function
was recently noticed by Tamm et al.~\cite{henn} in reply to an evaluation of a
one-loop response function in terms of dressed propagators for the electro gas
in metals and semiconductors\cite{schoene}. 

For self-consistent approximation schemes a general recipe for
calculating the in-medium coupling of the external potential to dressed
nucleons is known \cite{kadBaym,blarip}. 
The in-medium  vertex describing the coupling of the external perturbation to
the nucleons nucleons is given by the solution of the BS equation
(Fig. \ref{bsfig}), where $K$ denotes the particle-hole irreducible
kernel. The kernel $K$  of the BS equation 
should be taken consistently with the  chosen expression for the
self-energy. It is given by  the  functional derivative of the
self-energy with respect to the dressed Green's function 
\cite{kadBaym,blarip} $K={\delta \Sigma}/{\delta G}$.

\begin{figure}
\centering
\includegraphics*[width=0.8\textwidth]{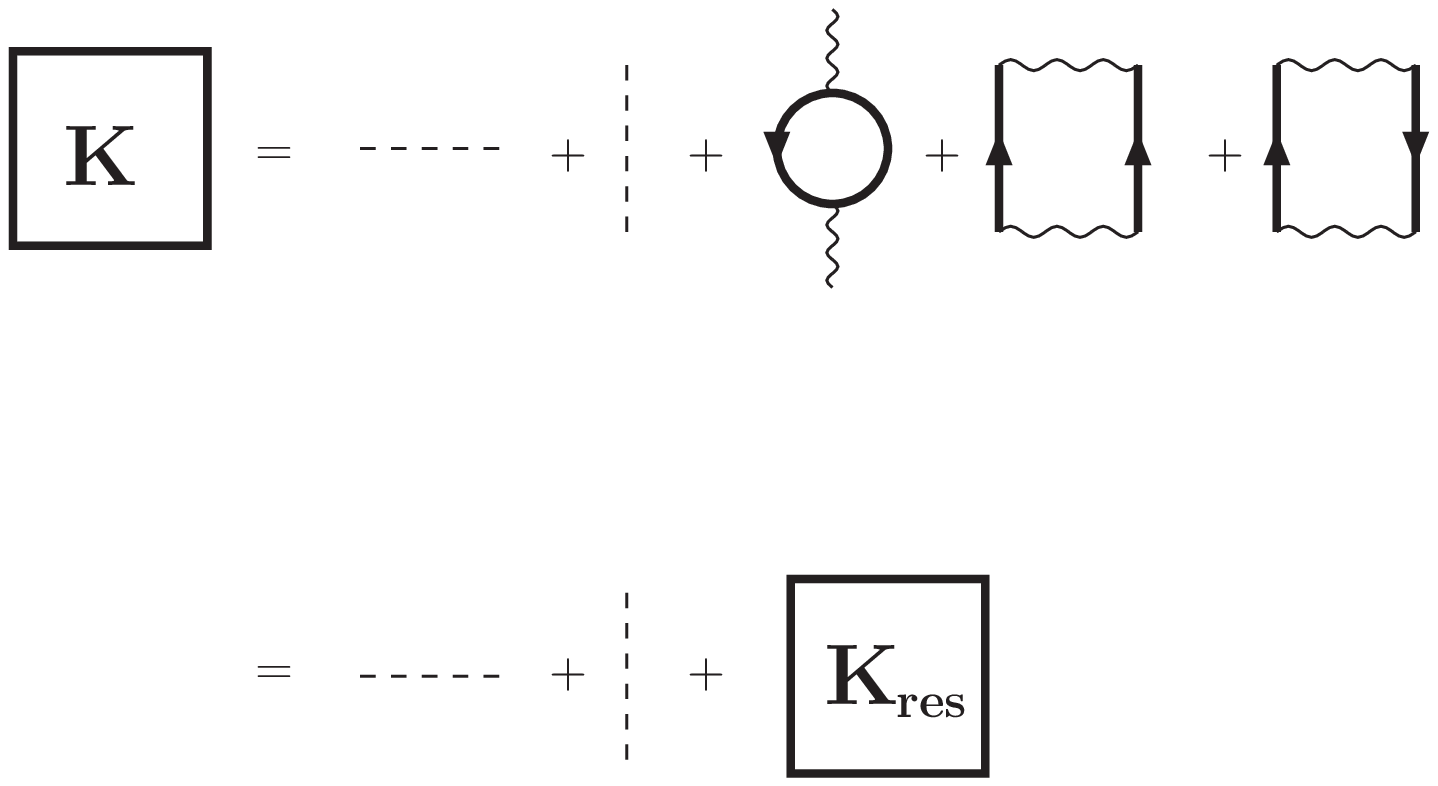}
\parbox{14cm}{\it\caption{The kernel of the  Bethe-Salpeter equation
corresponding to the self-energy in Fig. \ref{selffig}  containing a mean field
and a residual interaction. }
\label{kernelfig}}
\end{figure}

The resulting  kernel of the BS equation contains the usual mean-field
interaction (direct and exchange terms, see first and second term in the
representation of $K$ displayed in Fig. \ref{kernelfig}) and additional diagrams
which are due to the contributions of the residual interaction terms in the
self-energy and are collected as $K_{res}$ in  Fig. \ref{kernelfig}. 

\begin{figure}
\centering
\includegraphics*[width=0.5\textwidth]{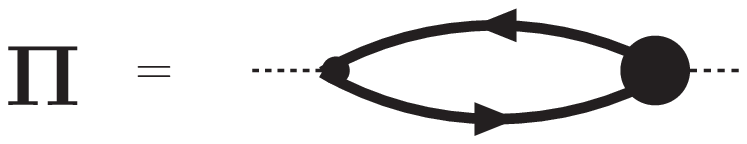}
\parbox{14cm}{\it\caption{The polarization function expressed using the dressed
  vertex for the coupling of the external current to the dressed nucleon.}
\label{polver}}
\end{figure}

Using the dressed vertex obtained as a solution of the BS equation the 
response function in the correlated medium can be obtained from the
diagram in Fig. \ref{polver}. Only one vertex in the loop includes in-medium
modifications in order to avoid double counting.

\begin{figure}
\centering
\includegraphics*[width=0.7\textwidth]{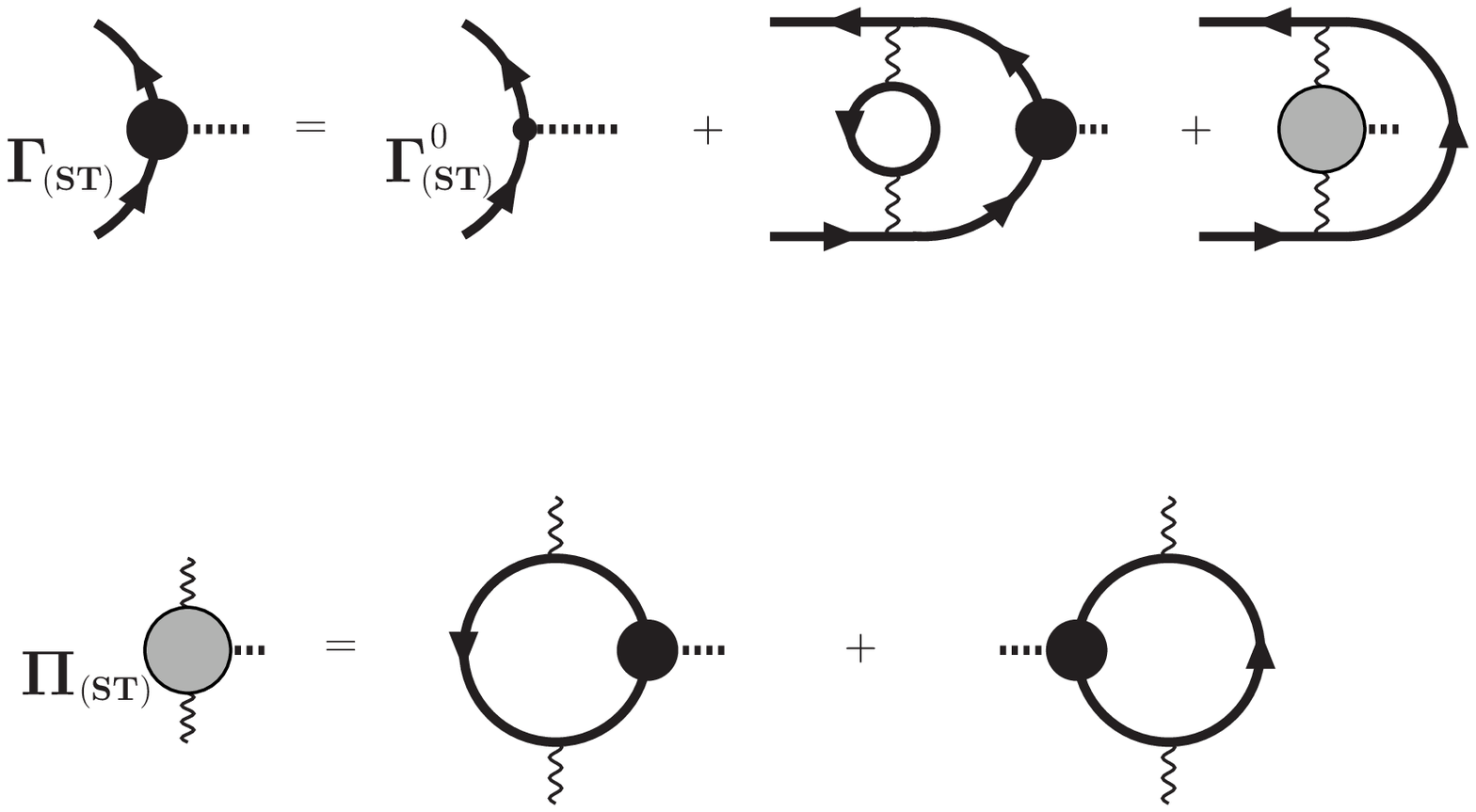}
\parbox{14cm}{\it\caption{The Bethe-Salpeter equation with contributions only
from the  residual interaction (Eqs. \ref{glg}, \ref{gpm}, and \ref{polext2})}
\label{bsfig2}}
\end{figure}

The three-point  Green's function for the coupling of the nucleon to
an external field with 
momentum ${\bf q}$ and energy $\Omega$ is
\begin{equation}
G_{(ST)}(x_1,t_1;x_2,t_2;{\bf q},\Omega)=-\int d^3x\, dt\, \exp(-i{\bf q x
  }+i\Omega t) \langle{\cal T} \Psi(x_1,t_1)\rho^{(ST)}(x,t)
\Psi^\dagger(x_2,t_2) \rangle
  \, ,
\end{equation}
where $\Psi^\dagger, \Psi$ are the field creation and annihilation
operators, $\rho^{(ST)}(x,t)=\Psi^\dagger(x,t)\Gamma_{(ST)}^0\Psi(x,t)$ denotes
the bare coupling to the external field with the spin-isospin operators
$\Gamma_{(ST)}^0=1, \ \sigma_3, \ \tau_3, \ \tau_3\sigma_3$ for the response
functions denoted by spin $S$ and isospin $T$ equal to $ST=00, \ 10, \
01, \ 11$, respectively. The operator ${\cal T}$ in this equation is the usual
operator for the time
ordering on the real time contour \cite{keldysh}. 

This three-point Green's function for
the dressed coupling of the external field to the fermions in medium has a
complicated analytical structure and depends on the incoming  
momentum ${\bf q}$ and energy $\Omega$ \cite{evans}. 
Depending on the ordering of the times of the fermion operators one
can define the smaller (larger) Green's functions
$G_{(ST)}^{<(>)}(x_1,t_1;x_2,t_2;{\bf q}, \Omega)$ and also the  retarded or
advanced ones.
In the momentum representation the three-point Green's function depends on
the momentum ${\bf p}$ and energy $\omega$ of the
incoming fermion and the  momentum is ${\bf q}+{\bf p}$ and  
energy $\omega+\Omega$ of the outgoing fermion. 
 We write the smaller  (larger) Green's functions
\begin{equation}
G_{{(ST)}}^{<(>)}({\bf q}+{\bf p},\omega+\Omega;{\bf p},\omega)
\end{equation}
and denote the retarded (advanced) Green's functions by
\begin{equation}
G_{{(ST)}}^{r(a)}({\bf q}+{\bf p},\omega+\Omega;{\bf p},\omega) \ .
\end{equation}
The response function can be expressed using this three-point Green's
function (Fig. \ref{polver})
\begin{equation}
\Pi_{{(ST)}}^{r}({\bf q},\Omega)=-i Tr \int \frac{d^3p d\omega}{(2\pi)^4}
\Gamma_{(ST)}^0 G_{(ST)}^{<}({\bf p}+{\bf q},\omega+\Omega;{\bf p},\omega) \ .
\label{eq:respons}
\end{equation}

The three-point Green's functions $G_{{(ST)}}$ can be written in terms of
the in-medium  (dressed) vertex $\Gamma_{(ST)}$ 
describing the in-medium coupling to the external perturbation 
\begin{equation}
\label{gdlg}
G_{{(ST)}}^{r(a)}({\bf q}+{\bf p},\omega+\Omega;{\bf p},\omega)=
G^{r(a)}({\bf q}+{\bf p},\omega+\Omega)\Gamma^{r(a)}_{(ST)}({\bf q}+{\bf
  p},\omega+\Omega;{\bf p},\omega)
G^{r(a)}({\bf p},\omega)
\end{equation}
and
\begin{eqnarray}
\label{gdpm}
\lefteqn{G_{{(ST)}}^{<(>)}({\bf q}+{\bf p},\omega+\Omega;{\bf p},\omega)
=}\hspace{3cm}
\nonumber\\
&&G^{r}({\bf q}+{\bf p},\omega+\Omega)\Gamma^{r}_{(ST)}({\bf q}+{\bf
  p},\omega+\Omega;{\bf p},\omega)
G^{<(>)}({\bf p},\omega) \nonumber \\
&&+ G^{r}({\bf q}+{\bf p},\omega+\Omega)\Gamma^{<(>)}_{(ST)}({\bf q}+{\bf
  p},\omega+\Omega;{\bf p},\omega)
G^{a}({\bf p},\omega) \nonumber \\
&&+ G^{<(>)}({\bf q}+{\bf p},\omega+\Omega)\Gamma^{a}_{(ST)}({\bf q}+{\bf
  p},\omega+\Omega;{\bf p},\omega)
G^{a}({\bf p},\omega) \ .
\end{eqnarray}
The dressed vertex $\Gamma_{(ST)}$ for the coupling of 
the external field to the
nucleon is the solution of the Bethe-Salpeter equation displayed in 
Fig. \ref{bsfig}.

As mentioned before the
  kernel of this BS equation has contributions from the mean-field 
and from the residual interaction. In this work we are interested in
the role of the correlations going beyond the mean-field, which are described by
means of the residual interaction discussed above. Therefore we will also
consider the effects which are due to the mean field and the
residual interactions in the kernel of the Bethe-Salpeter in separate
steps. In a first step we will concentrate on the effects of the residual
interaction and determine  a response function $\Pi_{{(ST)}
  \ res}$ which includes vertex correction from the residual interaction 
($K_{res}$ in Fig. \ref{kernelfig}), only. The way to evaluate $\Pi_{{(ST)}
  \ res}$ will be discussed below.

\begin{figure}
\centering
\includegraphics*[width=0.5\textwidth]{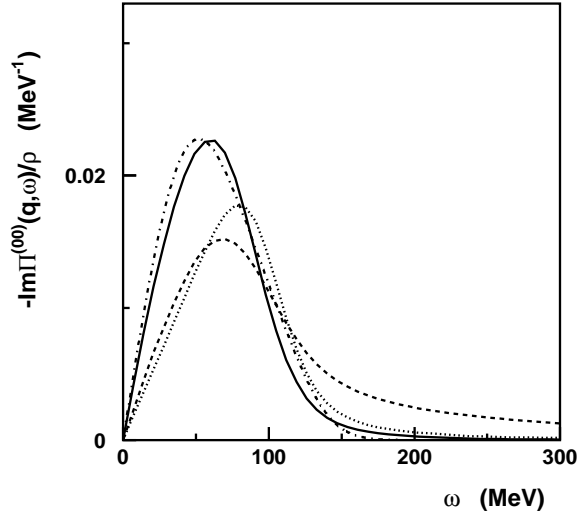}
\parbox{14cm}{\it\caption{The imaginary part of the response function as
function of the energy for the momentum $q=220$ MeV, at  normal nuclear density
and temperature $15$ MeV, without including the RPA modifications of the
response. The dashed-dotted  line is the response function for the free Fermi
gas with the same effective mass for the nucleons as in the correlated matter.
The solid line denotes the response function including the dressing of the
propagators and the vertex corrections (the density response). The dashed line
is the naive one-loop response function  with dressed propagators, but without
vertex corrections. The dotted line is the response including  the dressing of
the propagators and the vertex corrections from the  third diagram on the right
hand side in Fig. \ref{kernelfig} (spin-isospin channels $01$, $10$, and
$11$).}
\label{polirr}}
\end{figure}

 The modifications of the response function which are due to the mean-field
interaction can be taken into account by a solution of a separate
BS equation with a kernel including only this mean-field interaction
and the zero order vertex being the vertex dressed by the residual
interaction. In the following we approximate the solution of this second
BS equation with the mean-field kernel using Landau parameters for the 
mean-field interaction. E.g. for the density response function we have
\begin{equation}
\label{LandauRPA}
\Pi^r_{(00)}(p,\omega)
=\frac{\Pi^r_{{(00)}\ res}(p,\omega)}{1-f_0\Pi^r_{{(00)} \ res}(p,\omega)} 
\end{equation}
and analogously for other channels. 
The Landau parameters are calculated using the modified Gogny
mean-field with parameters from Table \ref{table}. There is 
little change of these parameters when comparing with the Landau
parameters obtained from the original Gogny interaction.
Only in the scalar channel the interaction gets less attractive.

So we now turn to the effects of the residual interaction in the response
function. In the real-time representation  the Bethe-Salpeter equations
including only the residual interaction in the kernel  take 
the form
\begin{eqnarray}
\label{glg}
\lefteqn{\Gamma^{<(>)}_{(ST)}({\bf p}+{\bf q},\omega+\Omega;{\bf p},\omega)=}
\hspace{1cm}\nonumber \\
&&i Tr \int \frac{d^3 p_1 d\omega_1 }{(2\pi)^4} \left( V^2({\bf p}-{\bf p_1})
\Pi^{<(>)}({\bf p}-{\bf p_1},\omega-\omega_1) \right. 
G_{(ST)}^{<(>)}
({\bf p_1}+{\bf q},\omega_1+\Omega;{\bf p_1},\omega_1) \nonumber \\
&&\left.  + V({\bf p_1})V({\bf p_1}+{\bf q})
\widetilde{\Pi}_{(ST)}^{<(>)}({\bf p_1}+{\bf q},\omega_1+\Omega;{\bf p_1},\omega_1)
G^{<(>)}({\bf p}-{\bf p_1},\omega-\omega_1) \right)
\end{eqnarray}
and
\begin{eqnarray}
\label{gpm}
\lefteqn{\Gamma^{r(a)}_{{(ST)}}({\bf p}+{\bf q},\omega+\Omega;{\bf p},\omega)=
1+ i Tr \int \frac{d^3 p_1 d\omega_1 }{(2\pi)^4}\Bigl(}
\hspace{1cm}\nonumber \\
&& V^2({\bf p}-{\bf p_1})
\Pi^{<}({\bf p}-{\bf p_1},\omega-\omega_1)  G_{(ST)}^{r(a)}
({\bf p_1}+{\bf q},\omega_1+\Omega;{\bf p_1},\omega_1) \nonumber \\
&&+V({\bf p_1})V({\bf p_1}+{\bf q})
\widetilde{\Pi}_{(ST)}^{<}({\bf p_1}+{\bf q},\omega_1+\Omega;{\bf p_1},\omega_1)
G^{r(a)}({\bf p}-{\bf p_1},\omega-\omega_1) \nonumber \\
&&+  V^2({\bf p}-{\bf p_1}) \Pi^{r(a)}({\bf p}-{\bf p_1},\omega-\omega_1)
G_{(ST)}^{>}
({\bf p_1}+{\bf q},\omega_1+\Omega;{\bf p_1},\omega_1)  \nonumber \\
&& +V({\bf p_1})V({\bf p_1}+{\bf q})
\widetilde{\Pi}_{(ST)}^{r(a)}({\bf p_1}+{\bf q},\omega_1+\Omega;{\bf p_1},\omega_1)
G^{>}({\bf p}-{\bf p_1},\omega-\omega_1) \Bigr) \ .
\end{eqnarray}
The terms in these equations containing the one-loop polarization function
$\Pi^{<(>)}$ or  $\Pi^{r(a)}$ of Eq.(\ref{polsimp}) and Eq.(\ref{disppol}),
respectively correspond to the second diagram on the right-hand side of the
first line in Fig. \ref{bsfig2}. This is the lowest order contribution of the
so-called induced interaction \cite{pico2,babu,pico3,friman} to the response
function and therefore we will refer to it as the induced interaction term in
the discussion below. The other terms refer to vertex corrections, which
represented by the third diagram on the right-hand side of the
first line in Fig. \ref{polext2}. They contain a three-point function which is
displayed in the second line of Fig. \ref{bsfig2} and is defined by  
\begin{eqnarray}
\label{polext1}
\lefteqn{\widetilde{\Pi}_{(ST)}^{<(>)}({\bf q}+{\bf
  p},\omega+\Omega;{\bf p},\omega)=-  i Tr 
  \int\frac{d^3p_1 d\omega_1}{(2\pi)^4}
\Bigl(}\hspace{1cm}\nonumber \\ &&G_{(ST)}^{<(>)}({\bf q}+{\bf
  p}+{\bf p_1},\omega+\Omega+\omega_1;{\bf p}+{\bf
  p_1},\omega+\omega_1)G^{>(<)}({\bf p_1},\omega_1)\nonumber \\
 &&
+ G^{<(>)}({\bf p}+{\bf p_1},\omega+\omega_1)G_{(ST)}^{>(<)}
({\bf p_1},\omega_1;
{\bf p_1}-{\bf q},\omega_1-\Omega) 
\Bigr) 
\end{eqnarray}
and
\begin{eqnarray}
\label{polext2}
\lefteqn{\widetilde{\Pi}_{(ST)}^{r(a)}({\bf q}+{\bf
  p},\omega+\Omega;p,\omega)=- i Tr   \int\frac{d^3p_1 d\omega_1}{(2\pi)^4}
\Bigl(}\hspace{1cm}\nonumber \\ && G_{(ST)}^{r(a)}({\bf q}+{\bf
  p}+{\bf p_1},\omega+\Omega+\omega_1;{\bf p}+{\bf
  p_1},\omega+\omega_1) G^{<(>)}({\bf p_1},\omega_1) \nonumber \\
&&+ G^{r(a)}({\bf p}+{\bf p_1},\omega+\omega_1)
G_{(ST)}^{<(>)}({\bf p_1},\omega_1;
{\bf p_1}-{\bf q},\omega_1-\Omega)
\nonumber \\
&&+ G_{(ST)}^{<(>)}({\bf q}+{\bf
  p}+{\bf p_1},\omega+\Omega+\omega_1;{\bf p}+{\bf
  p_1},\omega+\omega_1)G^{r(a)}({\bf p_1},\omega_1)
\nonumber \\ &&
+ G^{<(>)}({\bf p}+{\bf p_1},\omega+\omega_1)
G_{(ST)}^{r(a)}({\bf p_1},\omega_1;
{\bf p_1}-{\bf q},\omega_1-\Omega) 
\Bigr) \ .
\end{eqnarray}
For a scalar residual interaction the contributions of $\widetilde{\Pi}_{(ST)}$ 
are nonzero only for the density response, i.e. $S=T=0$.

\section{Results and discussion}

The numerical solution of the equations for the in-medium vertex is
exorbitantly difficult \cite{bonitz,sedr,faleev,pbpl}. This is mainly due to the
complex structure of the spectral functions. 
Therefore in the following we present results only at finite
temperature and for a relatively large value of the momentum transfer $q$. 
In this case the spectral functions are relatively
smooth and therefore easier to handle in numerical calculations.
Eqs. 
(\ref{gdlg}), (\ref{gdpm}), (\ref{glg})-(\ref{polext2}) 
are solved by iteration
 for each given $q$ and $\Omega$,
using
the Green's functions $G$ dressed by the self-energy (\ref{self}).
Using Eq.(\ref{eq:respons}) we can then calculate the response function
$\Pi^r_{{(ST)}\ res}$ which accounts for the effects of the 
residual interaction.

In Fig. \ref{polirr} we show the results for this polarization function
with vertex corrections $\Pi_{res}$ for $q=220$MeV. The results for the
imaginary part of the response function originating from the naive one-loop 
polarization (\ref{polsimp}) calculated with dressed propagators are represented
by the dashed line. As compared to the Hartree-Fock response function
(dashed-dotted line) this one-loop calculation with dressed propagators yields a
significant tail at large excitation energies. As the dressed propagators include
effects of two-particle one-hole and two-hole one-particle contributions to the
propagation of particle and holes, the dressed response functions
accounts for admixtures of two-particle two-hole contributions to the response
functions. Therefore one may interprete this high-energy tail to describe a
shift of the excitation strength to higher energies due to the admixture of
these two-particle two-hole contributions.

If, however, we also account for the vertex corrections which are due to the
residual interaction, we obtain the response functions represented by the solid
line in the case of the density response and the dotted line in the case of the
response for the other spin-isospin channels. One can see that in all cases the
high-energy tail obtained in the simple one-loop result is compensated by the
vertex corrections. This means that the induced interaction term, which for our
scalar residual interaction is present in all spin-isospin channels, is
responsible for this cancellation at high energies. 
The difference between the density response $\Pi_{(00)\ res}$ 
and the response  in other spin-isospin channels is due to the sub-leading vertex
corrections represented by the second and third graph in Fig.~\ref{kernelfig}
for $K_{res}$. These vertex corrections, which are specific for the density
response, lead to an enhancement of the imaginary part of $\Pi_{(00)\ res}$ at
small energies, which makes the final result look rather similar to the free
response function without any corrections of propagator and vertex due to the
residual interaction.

\begin{figure}
\centering
\includegraphics*[width=0.7\textwidth]{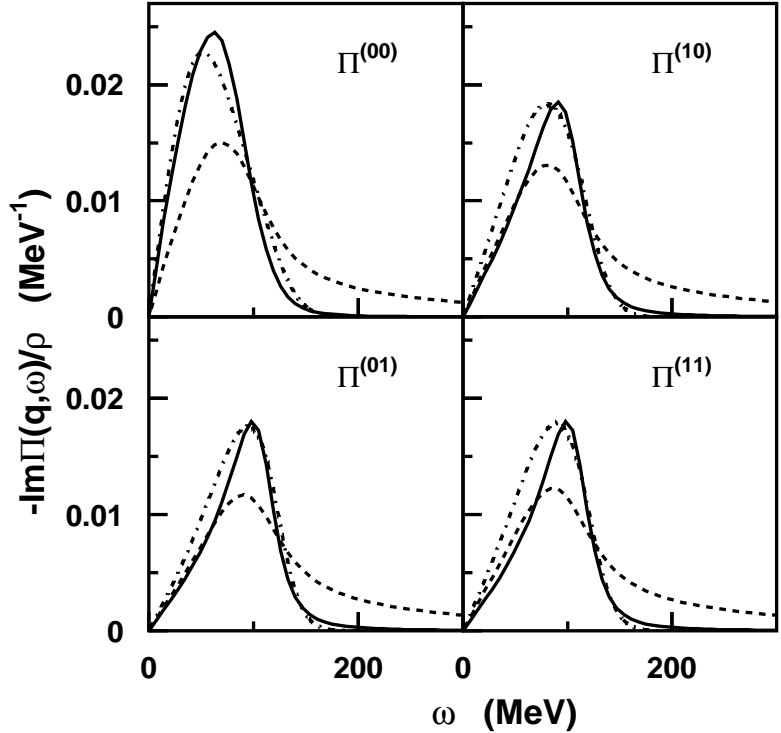}
\parbox{14cm}{\it\caption{The imaginary part of the response functions in
different channels as function of energy for the momentum $q=220$ MeV, at
normal nuclear density and temperature $15$ MeV. The dashed-dotted  line is the
response function for the Gogny interaction in the Landau parameter
approximation.The solid line denotes the response function including the
dressing of the propagators and the vertex corrections.  The dashed line is the
naive one-loop response function  with dressed propagators. The response
function for the system with residual interaction include the modification of
the response due to the modified Gogny mean-field, taken in Landau parameter
approximation.}
\label{polfig}}
\end{figure}
The free response and the consistently calculated response functions
in the correlated system should fulfill several sum rules.
For the scalar residual interaction  the $\omega$-sum rule takes the
simple form
\begin{equation}
\label{sumreq}
-\int_{-\infty}^{\infty} \frac{\omega d \omega}{2\pi}{\rm Im}\Pi^r_{(ST)}
({\bf q},\omega)= \rho \frac{{\bf q}^2}{2m} \ ,
\end{equation}
in all the spin isospin channels ${(ST)}$.
The self-energy takes into account also the mean-field
interaction, so the sum rule is only approximate and since the response
$\Pi_{res}$ does not include the RPA corrections on the right hand
side of (\ref{sumreq}) 
we substitute the free mass with the effective mass. Such a modified 
sum
rule is fulfilled to within a few  percent by the response functions
including vertex corrections $\Pi_{(ST) \ res}$,
 it is severely violated by the naive
one-loop response function (\ref{polsimp}).
 The mean-field interaction contains spin
and isospin dependent terms, and the sum rule including the mean-field
part of the interaction could include a possible RPA enhancement factor
\cite{liparini}
besides the free Fermi gas sum rule (\ref{sumreq}). When restricting
the RPA response to the Landau parameter form (\ref{LandauRPA}) the 
$\omega$-sum rule has the same form as in the Free Fermi gas
(\ref{sumreq})
but with the corresponding effective mass instead of the free nucleon mass.
 We find that
 the vertex corrections due to the first diagram 
 for  the kernel $K_{res}$ (Fig. \ref{bsfig2}), the so-called induced
interaction terms, are the most important ones
to bring the response function close to the one for the Free Fermi
case and  restore the $\omega$-sum rule.

Adding the Hartree-Fock terms in the self-energy modifies the kernel
of the equation for the dressed vertex. 
As explained above we take  the RPA sum 
 into account by means of the Landau parameter for
the mean-field part of the interaction.
The resulting response functions in different channels are shown in
Fig. \ref{polfig}.  We plot also the response function for a Fermi
liquid, where the Landau parameters and the effective mass
 are given by the original Gogny interaction.
For the density response the result is very close
to the response of a Fermi liquid. 
In all the channels the naive one-loop polarization 
with dressed propagators gives a incorrect description,
with long tail at large energies. In fact the Lindhard function, i.e.
the one-loop polarization with HF propagators, gives a much better
description of the response function \cite{pines}, similar to the one including
full dressed propagators and vertices.
In the spin isospin response some
difference to the response of a Fermi liquid is observed, which could
already be seen in Fig \ref{polirr}. However the constraint of the
$\omega$-sum rule makes the response in the correlated systems
to lie close to the Fermi liquid one also in the nonzero spin and/or
isospin channels; the overall shape of $\mbox{Im }\Pi_{(ST)}$ is similar to the RPA
response function.

\subsection{Neutron matter}

The description of weak processes in dense nuclear matter is very
important for modeling supernovae explosions and the cooling 
of neutron stars.
In a hot and dense medium neutrinos have a short mean free path and
they are effectively trapped inside the proto-neutron star.
The calculation of the mean free path involves nuclear correlation
effects. The relevant hadronic part of the cross section can be
factorized in the form of the density and spin response in matter.
In this section we present a calculation the response functions in
pure neutron matter.

As for the symmetric nuclear matter, the  mean-field interaction has to be
modified in order to take into account additional contributions from the
residual interaction. In this first exploratory work we opt for a 
parameterization which is different in pure neutron matter and in symmetric 
nuclear matter. The modifications of the mean field interaction are listed in 
the third colum of parameters in Table \ref{table}. In this way we can
reproduce the same Fermi energy and similar effective mass  as  given by the
original Gogny interaction for a range of densities between $0.4\, \rho_0$ and
$\rho_0$.  At the same time the Landau parameters are not modified drastically
from their value corresponding  to  the original  Gogny parameters displayed in
the first column of Table \ref{table}.

\begin{figure}
\centering
\includegraphics*[width=0.5\textwidth]{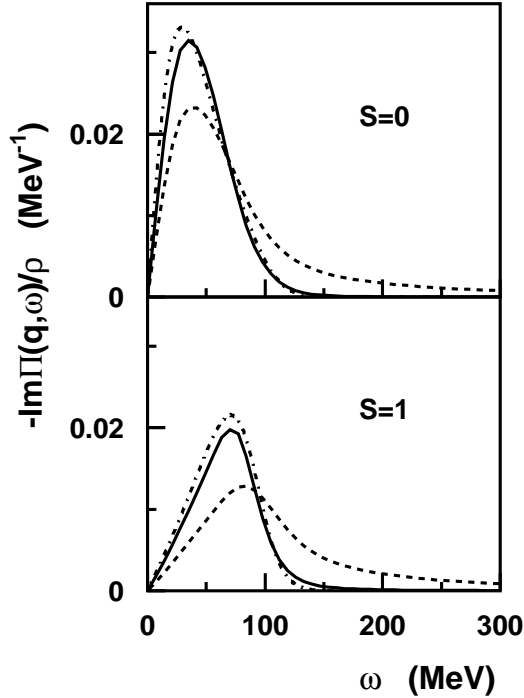}
\parbox{14cm}{\it\caption{The imaginary part of the response functions in
different channels as function of energy for the momentum $q=220$MeV, in
neutron matter at normal density. Symbols are the same as in Fig.
\ref{polfig}.}
\label{polne}}
\end{figure}

The formulas for the density and spin response in neutron matter can
be written in the same way as outlined in section \ref{respsect}.
 We find that
for the density response the whole kernel $K_{res}$ displayed in
Fig.~\ref{bsfig2} must be considered in the 
BS equation, while for the spin response only the first graph in the
kernel $K_{res}$, the induced interaction term, is nonzero 
($\widetilde{\Pi}_{{(S)}}=0$ for $S=1$).

The results are very similar to what we found for the symmetric
nuclear matter.  When
both propagator and vertex modification in the medium are taken into
account the response function in the correlated system is very similar 
to the one obtained in the Fermi liquid theory
(Fig \ref{polne}). It is not surprising,
since the $\omega$-sum rule has the same form as in the
noninteracting system.
The naive one-loop response function with dressed propagators 
cannot be trusted and violates the sum rule.

\subsection{Multi-pair contributions to the response function}

Both for the symmetric and pure neutron matter we find that the
response function in a correlated system is very close to 
response function in free Fermi gas, or when the mean field is taken
into account the response function is similar as in the Fermi liquid
theory. This means that the cancellation of propagator dressing and vertex
correction effectively drives the response of the system to the 
response given by the excitation of a single particle-hole pair. 

However, this result is not general. This cancellation is
due to the particular form of
the residual interaction, which we have considered to be scalar in spin and 
isospin. This also leads to
the simple form (\ref{sumreq}) of the $\omega$-sum rule in all the channels.

\begin{figure}
\centering
\includegraphics*[width=0.7\textwidth]{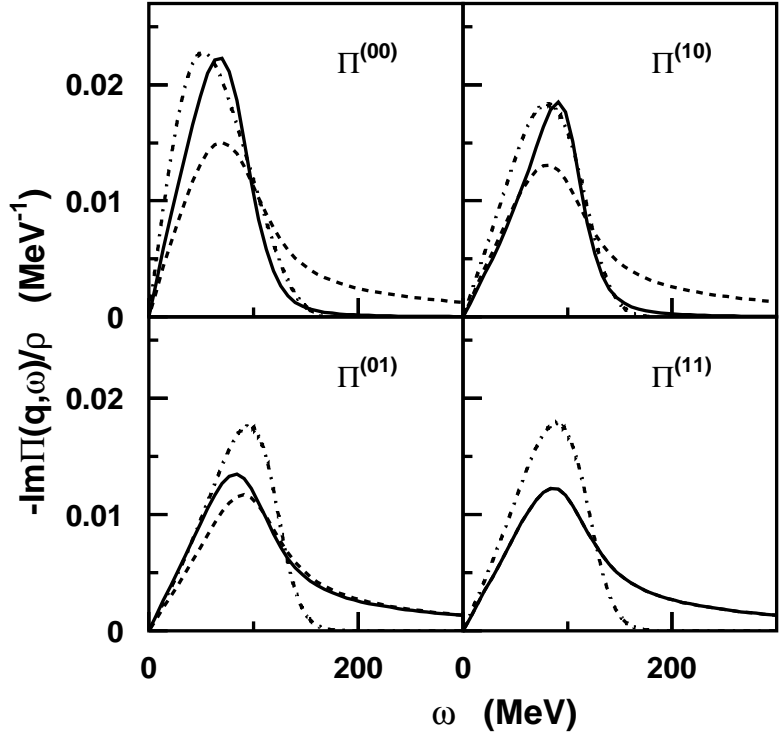}
\parbox{14cm}{\it\caption{The imaginary part of the response functions in
different channels as function of energy and $q=220$MeV  for the isospin
dependent residual interaction (\ref{newres}). Symbols are the same as in Fig.
\ref{polfig}. The $ST=11$ response function with vertex and propagator
dressing  is the same as the naive one-loop result with only propagator
dressing. }
\label{poliso}}
\end{figure}

Therefore, for the discussion in this section we modify the residual 
interaction and assume it to be isospin dependent in the form
\begin{equation}
\label{newres}
V_{res}({\bf r_1}-{\bf r_2})P_\tau
=V_0 e^{-({\bf r_1}-{\bf r_2})^2/2\eta^2}P_\tau
\ .
\end{equation}
In symmetric nuclear matter the single-particle self-energy 
is the same as obtained for the scalar residual interaction.
Therefore the same modified Gogny parameterization of the mean field
interaction is used  as
in section \ref{residualmf}.
The kernel $K_{res}$ of the BS equation, however, is different than in the case
of a scalar residual interaction.
For the $ST=00$ channel all the three graphs for $K_{res}$  in
 Fig. \ref{kernelfig} 
contribute. However, for the response function in channels with isospin
$T=1$ the first diagram in $K_{res}$, the induced interaction term does not 
contribute. We have found previously that this induced interaction diagram in
  the vertex dressing is crucial for the suppression of the high-energy tail in
the response function. This led to the restoration of the 
$\omega$-sum rule and made  the response function 
  similar to the one in the free Fermi gas. 

So if this induced interaction contribution to the
  vertex dressing is absent (channels $ST=01, \ 11$) we expect a
  strong modification of the response function by the residual
  interaction.
The vertex and propagator dressing do no longer cancel. In fact,
in the channel $ST=11$ there are no
vertex corrections at all ($K_{res}=0$), and the response function has
the same form as a naive one-loop calculation with dressed propagators.

In Fig. \ref{poliso} the  response functions obtained with the isospin
dependent interaction (\ref{newres}) are compared to
the response functions from the Fermi liquid theory.
For $T=0$ channels the correlated response function is similar to the
one particle-one hole response function. On the other hand,
the isovector response is  closer to the naive
one-loop result.

For the $T=1$ channels the $\omega$-sum rule is different than in the
free Fermi gas.
The residual interaction in the Hamiltonian
gives a modification factor in the $\omega$- sum rule at finite
momentum $q$
\begin{eqnarray}
\label{sumr2}
\lefteqn{-\int_{-\infty}^{\infty} \frac{\omega d \omega}{2\pi}{\rm Im}\Pi^r_{(ST)\ res}({\bf q},\omega)=
\rho \frac{{\bf q}^2}{2m}}\hspace{2 cm} \nonumber \\ 
&&+ \frac{2}{3}\int d{\bf r}_1 d{\bf r}_2
V_{res}({\bf r_1}-{\bf r_2}) \tau_1\tau_2
\Psi^\dagger(r_1)\Psi^\dagger(r_2)\Psi(r_2)\Psi(r_1)\left(
e^{i{\bf q}({\bf r}_1-{\bf r}_2)}-1\right) \ .
\end{eqnarray}
This enhancement of the sum rule in the $T=1$ response is consistent
with observed long tail in the response function $\mbox{Im }\Pi$ at large
energies. A nonzero value of the imaginary part of the
response function at large energies is not kinematically
allowed by one particle-one hole configuration with on shell
propagation. Nonzero  contribution do appear due to the dressing of the
single-particle propagator by the self-energy from the residual
interaction. Such a dressed propagation involves nucleons which are put
off shell by the scattering on other particles in the medium. 
For the isospin dependent interaction and $T=1$ response
 these off-shell propagation
effects are not canceled by vertex corrections. In the case of the
residual interaction of the form (\ref{newres})  off-shell nucleons
couple in the same way as  free nucleons to  isovector potentials.

For a general residual interaction containing scalar, spin, and isospin
dependent terms,  we expect that the spin and isospin responses in a
correlated system  lie in between the naive one-loop result and the
Fermi liquid theory result. 

\subsection{Collective modes}

The response function may show a pronounced peak at a certain excitation
energy. This is a collective mode, which corresponds to the excitation of a
single collective state in the interacting system. Depending on the
spin-isospin character of the response these are the zero sound mode, spin or
isospin waves.  In nuclear physics the isovector
response is of particular importance \cite{braghin}.
In finite nuclei it shows up as the giant dipole resonance, which has  
extensively been studied. 

Within the Fermi Liquid theory a collective excitation at zero temperature is a
discrete peak in the imaginary part of the  response function. The state
corresponding to the collective excitation  cannot couple to the incoherent one
particle-one hole excitations. At finite temperature such a coupling is
possible, it can be calculated and the width of the collective state at finite
temperature is usually small. The collective state can acquire a finite width
(also at zero temperature) due to a coupling to multi pair configurations
\cite{pines}. The description of this damping of the collective states 
from such admixtures  goes beyond the usual Fermi liquid theory. In the
preceeding subsection we have seen that a isospin dependent residual
interaction can produce correlations in the response function which correspond
to the admixture of multi pair configurations.

\begin{figure}
\centering
\includegraphics*[width=0.5\textwidth]{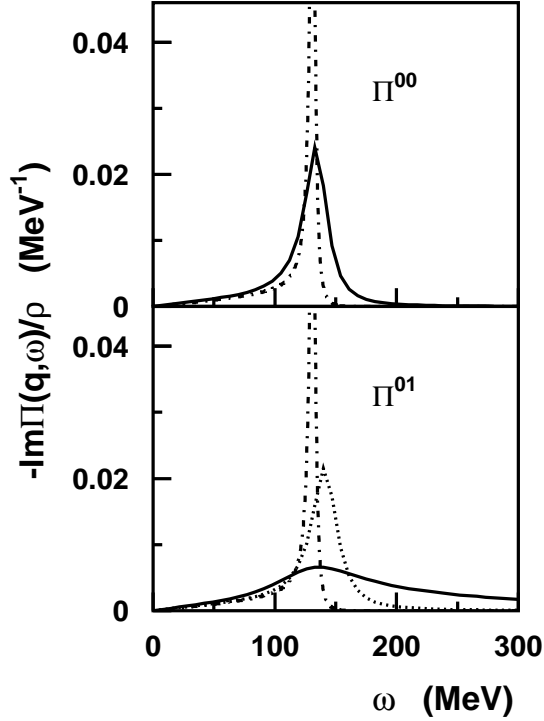}
\parbox{14cm}{\it\caption{The imaginary part of the response functions  as
function of energy and $q=220$MeV  for the isospin dependent residual
interaction (\ref{newres}) (solid lines) and for the scalar residual
interaction (\ref{vres}) (dotted line).  Dashed-dotted line are results for a
Fermi liquid at finite temperature. In order to get a collective mode for the
chosen momentum and temperature the Landau parameters are set by hand to $F_0,\
F_0^{'}=4$.}
\label{collfig}}
\end{figure}

For the chosen temperature and kinematics, however, the Hamiltonian considered
here does not lead to strong collective modes in any of the response functions.
To study the role of the multi pair configurations on the collective
modes we increase the value of the Landau parameters.
In Fig. \ref{collfig} we present the density response function assuming a Landau
parameter  
$F_0=4$. In this case the RPA response function shows a well defined
peak. 
The relatively high
temperature yields a collective zero sound mode with a finite width due to
the coupling to thermally excited one particle-one hole states.
The calculation including multi-pair correlations in the system from
the residual interaction does also show a collective state in the
density response. The position of this collective mode is almost at
the same place as for the Fermi liquid (Fig. \ref{collfig}).

The $ST=00$ response function with propagator and vertex
corrections  is  almost the same as in the free Fermi gas. 
The difference is that 
at high energy the response function $ \mbox{Im } \Pi_{ (00)\ res}$ is slightly
larger than the finite temperature Lindhard function.
 This causes the collective mode to 
have a larger width in the system with residual interactions. The
damping  of the zero
sound has two origins in a system interacting with a residual
interaction:
   a finite temperature width and a width due to the coupling
to multi-pair states.
 
In the lower panel of Fig. \ref{collfig} the isovector
response function is shown for the Landau parameter $F_0^{'}=4$. The Fermi
liquid theory predicts the presence of a well defined collective state.
The finite width is due of course to the finite temperature.
For the scalar residual interaction the response function in the
correlated system shows a collective state at similar energy.
It has a larger width due to the contribution of multi-pair
configurations, analogously as in the density response.

The isospin dependent residual interaction leads to a response
$\mbox{Im } \Pi_{(01) \ res}$ with a long tail at large energies. 
Due to the large contribution of these configurations the
width of the collective
states in the isovector response is very large. In fact, the 
collective mode disappears. The disappearance of the collective mode is
an extreme case  where the coupling to multi-pair state is not reduced
by  vertex corrections (special case of the interaction
(\ref{newres})). 

For a general interaction we  expect a whole range
of behavior depending on the energy of the collective state and on
the strength of isospin dependent terms in the residual interaction. 
The collective state would be generally broader than in the Fermi
liquid theory, due to the coupling to multi-pair configurations. In some
cases this coupling can lead to a disappearance of the collective state. 
The same phenomena are expected also for the spin wave 
collective
state in the presence of spin dependent residual interactions.

\section{Conclusions}

The aim of this paper has been a consistent study of correlation effects on the
response function going beyond the usual HF plus RPA approach. For that purpose
we consider a  mean-field interaction and a
residual interaction. This residual interaction generates contributions to the
self-energy of the nucleons, which describe the admixture of two-hole
one-particle and two-particle one-hole configurations to the single-particle
propagator.

The response function using these dressed propagators in a one-loop
approximation will in general violate the energy weighted sum
rule for the excitation function. These sum-rules are fulfilled only if the
response functions are calculated employing a consistent treatment of propagator
and vertex corrections following the recipe of Baym and Kadanoff
\cite{kadBaym,blarip}. A scheme for such a consistent treatment of correlation
effects in the response function of nuclear matter is outlined and numerical
results are presented for symmetric nuclear matter and pure neutron matter at
finite temperature.

Assuming a residual interaction of scalar-isoscalar form it turns out that the
effects originating from propagator
 corrections are to a large extent compensated
by vertex corrections in the response function for all spin-isospin channels. 
The induced interaction terms in particular are responsible for the 
compensation
of the correlation effects in the single-particle propagator.

If, however, a residual interaction with non-trivial spin isospin structure is
considered this cancellation of correlation effects is removed in specific
spin-isospin channels. Consequences for the damping of collective excitation
modes due to these admixtures of multi-particle multi-hole contributions are
discussed.

The present investigation employs residual interaction with a rather simple
spin-isospin structure. More realistic interaction models should be 
investigated  in extended kinematical regions of $q$ and $\Omega$
to obtain detailed information on the importance of correlation effects on the
nuclear response for the various excitation modes.

 {\bf Acknowledgments}
\vskip .3cm
We thank Armen Sedrakian for useful and interesting discussions.
P.B. would like to acknowledge the support  of the KBN
under Grant No. 2P03B05925 (Poland) and of the Humboldt Stiftung (Germany).


\begin{thebibliography}{99}

\bibitem{pines} D. Pines, P. Nozi\`eres, {\em The Theory of
Quantum Liquids Vol. I}, Benjamin, New York, 1966.

\bibitem{mf}
B.~Friman, O.~Maxwell, Astophys. J., {\bf 232} (1979) 541.

\bibitem{reddyne}
S.~Reddy, M.~Prakash, J.~Lattimer, J.~Pons, Phys. Rev., {\bf C59} (1998) 2888.

\bibitem{voskrn}
D.~Voskresensky, A.~Senatorov, Sov. Phys. JETP, {\bf 63} (1986) 885.

\bibitem{yamada}
S.~Yamada, H.~Toki, Phys. Rev., {\bf C61} (1999) 015803.

\bibitem{raffelt}
G.~Raffelt, D.~Seckel, Phys. Rev. Lett., {\bf 67} (1991) 2605.

\bibitem{sedr}
A.~Sedrakian, A.~Dieperink, Phys. Rev., {\bf D62} (2000) 083002.

\bibitem{haensel}
D.~Yakovlev, A.~Kaminker, O.~Gnedin, P.~Haensel, Phys. Rep., {\bf 354} (2001)
  1.

\bibitem{jerom1} J.~Margueron, J.~Navarro and P.~Blottiau, Phys. Rev. {\bf C70} 
(2004) 028801. 

\bibitem{pico1} 
S.-O.~B\"{a}ckmann and W.~Weise, in {\it Mesons in Nuclei,} edited by M.~Rho and
D.H.~Wilkinson (North-Holland, Amsterdam 1979) 1095. 

\bibitem{pico2}
W.H.~Dickhoff, A.~Faessler, J.~Meyer-ter-Vehn and H.~M\"{u}ther, Phys. Rev., 
{\bf C23} (1981) 1154.

\bibitem{Bozek:1997rv}
P.~Bo\.zek, Phys. Rev., {\bf C56} (1997) 1452.

\bibitem{Bozek:1998ro}
P.~Bo\.zek, in: K.~Morawetz, P.~{Lipavsk\'y}, V.~{\v{S}pi\v{c}ka} (Eds.), 5th
  Workshop on Nonequilibrium Physics at Short Time Scales, {Universit\"at}
  Rostock, Rostock, 1998.

\bibitem{eff}
M.~Effenberger, U.~Mosel, Phys. Rev., {\bf C60} (1999) 051901.

\bibitem{Cassing:2000ch}
W.~Cassing, S.~Juchem, Nucl. Phys., {\bf A677} (2000) 445.

\bibitem{Bertsch:1996ig}
G.~F. Bertsch, P.~Danielewicz, Phys. Lett., {\bf B367} (1996) 55.

\bibitem{Knoll:1996nz}
J.~Knoll, D.~N. Voskresensky, Annals Phys., {\bf 249} (1996) 532.


\bibitem{skyrme} T.H.R. Skyrme, Phil. Mag. {\bf 1} (1956) 1043.

\bibitem{gogny}  D. Gogny and R. Padjen, Nucl. Phys. {\bf A293} (1977) 365.

\bibitem{polrep} H. M\"uther and A. Polls, Prog. Part. and Nucl. Phys. 
{\bf 45} (2000) 243.

\bibitem{Bozek:2002em}
P.~Bo\.zek, Phys. Rev., {\bf C65} (2002) 054306.

\bibitem{Dewulf:2003nj}
Y.~Dewulf, W.~H. Dickhoff, D.~Van~Neck, E.~R. Stoddard, M.~Waroquier, Phys.
  Rev. Lett., {\bf 90} (2003) 152501.

\bibitem{Frick:2003sd}
T.~Frick, H.~M\"uther, Phys. Rev., {\bf C68} (2003) 034310.

\bibitem{rohe:2004}
T. Frick, Kh.S.A. Hassaneen, D. Rohe, and H. M\"uther, Phys. Rev., {\bf C70} 
(2004) 024309.

\bibitem{kadBaym}
G.~Baym, L.~Kadanoff, Phys. Rev., {\bf 124} (1961) 287.

\bibitem{blarip}
J.~Blaizot, G.~Ripka, {\em Quantum Theory Of Finite Systems}, MIT Press,
  Cambridge, 1986.

\bibitem{bonitz}
N.~Kwong, M.~Bonitz, Phys. Rev. Lett., {\bf 84} (2000) 1768.

\bibitem{faleev}
S.~Faleev, M.~Stockman, Phys. Rev., {\bf B66} (2002) 085318.

\bibitem{pbpl} P. Bo\.zek, Phys. Lett. {\bf B579} (2004) 309.

\bibitem{GNVS} C. Garc\'ia-Recio, J. Navarro, N. Van Giai,
L.L. Salcedo, Ann. Phys. {\bf 214} (1992) 293. 
 
\bibitem{depace} A. De Pace, Nucl. Phys. {\bf A635} (1998) 163.

\bibitem{jerom2} J.~Margueron, J.~Navarro, and N.~van Giai, Nucl. Phys. {\bf
A719} (2003) 169.

\bibitem{gognyd1p}  M. Farine et al., J. Phys. G {\bf 25} (1999) 25.

\bibitem{dan}P.~Danielewicz, Annals Phys., {\bf 152} (1984) 305.

\bibitem{keldysh} L.V. Keldysh, Zh. Eksp. Teor. Fiz. {\bf 47} (1964), 
1515.

\bibitem{bcapp}
P.~Bo\.zek and P.~Czerski,
Acta Phys.\ Polon.\ B {\bf 34} (2003) 2759.

\bibitem{oset} L. Alvarez-Ruso, P. Fernandez De Cordoba, E. Oset,
Nucl. Phys. {\bf A606} (1996) 407.

\bibitem{Lehr:2000ua}
J.~Lehr, M.~Effenberger, H.~Lenske, S.~Leupold, U.~Mosel, Phys. Lett., {\bf
  B483} (2000) 324.

\bibitem{henn}
D.~Tamme, R.~Schepe, K.~Henneberger, Phys. Rev. Lett., {\bf 83} (1999) 241.

\bibitem{schoene}
W.-D.~Schoene and A.G. Eguiluz, Phys. Rev. Lett., {\bf 81} (1998) 1662.

\bibitem{evans}
T.~S. Evans, Phys. Lett., {\bf B249} (1990) 286.

\bibitem{babu} S. Babu and G.E. Brown, Ann. of Phys. {\bf 78} (1973) 1.

\bibitem{pico3} W.H. Dickhoff and  H. M\"uther, Nucl. Phys.
{\bf A473} (1987) 394.

\bibitem{friman} A. Schwenk and B. Friman, Phys. Rev. Lett. {\bf 92} (2004) 
082501.

\bibitem{liparini} E. Lipparini and  S. Stringari, Phys. Rep. {\bf
  175} (1989) 103.

\bibitem{braghin} F.L.~Braghin, D.~Vautherin, and A.~Abada, Phys. Rev. {\bf C52}
(1995) 2504.






















\end{thebibliography}

\end{document}